\documentclass[twocolumn,showpacs,showkeys,superscriptaddress,amsmath,amssymb,citeautoscript,aps,10pt]{revtex4-1}

\usepackage{dcolumn}
\usepackage{amssymb}
 \usepackage{amsthm}
\usepackage[dvipsnames]{xcolor}
\usepackage{graphicx}
\usepackage[utf8]{inputenc}

\def \bsfA   {\mbox{\boldmath$\mathcal{A}$}}

\usepackage[hidelinks]{hyperref}

\makeatletter
\providecommand{\doi}[1]{%
  \begingroup
    \let\bibinfo\@secondoftwo
    \urlstyle{rm}%
    \href{http://dx.doi.org/#1}{%
      doi:\discretionary{}{}{}%
      \nolinkurl{#1}%
    }%
  \endgroup
}
\makeatother

\begin{document}

\title{Anomalous Hall and Nernst effects in a two-dimensional electron gas\\ with an anisotropic cubic Rashba spin-orbit interaction}
\author{A. Krzy\.zewska}
\affiliation{Faculty of Physics, Adam Mickiewicz University, 61-614 Pozna\'n, Poland}
\author{A. Dyrda\l{}}
\email{adyrdal@amu.edu.pl}
\affiliation{Faculty of Physics, Adam Mickiewicz University, 61-614 Pozna\'n, Poland}
\affiliation{Institut für Physik, Martin-Luther-Universität Halle–Wittenberg, 06099 Halle (Saale), Germany}

\begin{abstract}
The anomalous Hall and Nernst effects are considered theoretically within Matsubara-Green's function formalism. The effective Hamiltonian of a magnetized two-dimensional electron gas with cubic Rashba spin-orbit interaction may describe transport properties of electronic states at the interfaces or surfaces of perovskite oxides or another type of heterostructures that, due to symmetry, may be described by the same effective model. In the quasi-ballistic limit, both effects are determined by the topological (Fermi sea) contribution whereas the states at the Fermi level gives a negligibly small response. For a wide range of parameters describing the considered system, the anomalous Nernst conductivity reveals a change of the sign before the magnetic phase transition. 
\end{abstract}



\maketitle



\section{Introduction}

Spin-orbit interaction is the origin of various phases and phenomena observed in the physics of solid-state providing pure electric control of the spin degree of freedom~\cite{Soumyanarayanan2016Nov}. Nowadays, the spin-orbit driven transport phenomena, such as anomalous and spin Hall effects, has become a fundamental tool for generation spin accumulation and detection of spin currents and topological character of quasiparticles states. Moreover, pure electrical control of the spin degree of freedom is one of the crucial ideas of spintronics, according to which spin-based electronics should provide smaller, cheaper and faster electronic devices  with high functionality (e.g., data storage and logic operations in one material) and low energy consumption at room temperatures~\cite{Awschalom2007Oct,Bader2010Jul,Sinova2012Apr,Awschalom2007Mar,Fabian2007Aug}.

It is known that spin-orbit interaction strongly depends on the type of impurities and the crystallographic potential of the host material and is especially enhanced in low dimensional systems. In such a case the space inversion symmetry is broken at the surfaces or interfaces what results in an additional component of the spin-orbit interaction, called the Bychkov-Rashba interaction~\cite{Bychkov1984Jan,Winkler2003,Engel2007Dec,Bercioux2015Sep}. This type of spin-orbit interaction, resulting from structural inversion asymmetry, has been described initially in the context of a two-dimensional electron gas forming at the interfaces of semiconductor heterostructures~\cite{Winkler2003}.
For symmetry reasons, Rashba Hamiltonian is odd in quasiparticle momentum what leads in the simplest approximation to the well known $\mathbf{k}$-linear dependence. However, in various 2D systems,  terms with a cubic momentum dependence play also an important role. The so-called cubic Rashba interaction is responsible, e.g., for spin and transport properties of a two-dimensional hole gas formed at the interfaces of  III-V semiconductor heterostructures~\cite{Luttinger1955Feb,Luttinger1956May,Winkler2000Aug,Winkler2008Oct,Liu2008Mar,Moriya2014Aug}. Recently, the cubic character of Rashba interaction has also been found in a two-dimensional electron gas at the perovskite oxides surfaces and interfaces such as LaAlO$_3$/SrTiO$_3$ (LAO/STO)~\cite{Kim2013Jun,Shanavas2016Jan,Zhou2015Jun,Seibold2017Dec}. 

The thin films or heterointerfaces of perovskite oxides are a diverse group of materials with intriguing aspects of fundamental physics. 
For instance, the interfaces of insulating nonmagnetic oxide perovskites reveal interesting physical properties such as two-dimensional metallic conductivity, large negative magnetoresistance, metal-insulator transition, low-temperature superconductivity, and ferromagnetism as well as their coexistence~\cite{Brinkman2007Jun,Reyren2007Aug,Michaeli2012Mar}. Moreover, experimental data indicate strong spin-to-charge interconversion effects governed by spin-orbit coupling~\cite{Lesne2016Aug,Chauleau2016Oct,Song2017Mar,Wang2017Dec}.

\begin{figure*}
\centering
	\includegraphics[width=1\textwidth]{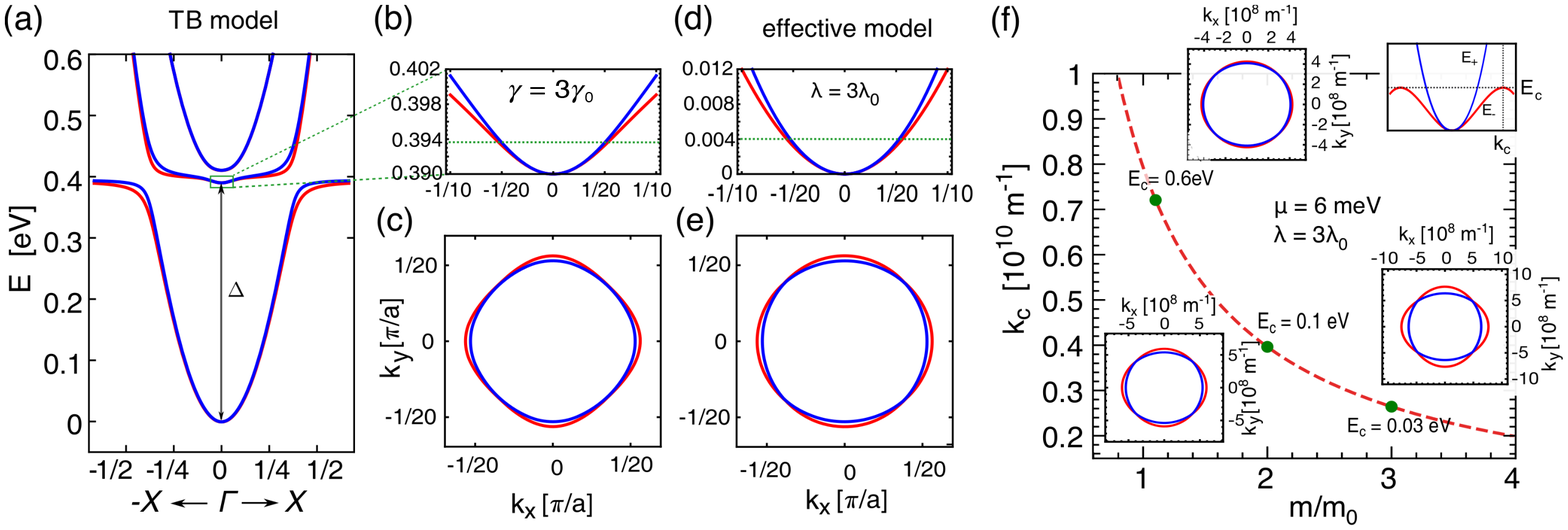}
	\caption{Band structure of the electronic states at the interface of  LAO/STO obtained based on TB model (a)-(c) and effective model for close vicinity of the $\Gamma$ point (d),(e). Figures (c) and (d) show constant energy contours at 4 meV above the bands minima in TB and effective model respectively. The anisotropy of energy spectrum in the $\mathbf{k}$-space is seen in both models. Figure (f) presents the cut-off wavevector as a function of the effective mass, and Fermi contours obtained for selected values of effective mass. The TB model and parameters such as lattice constant $a = 4.05\,\AA$, hopping integrals: $t_{1} = 0.227$ eV, $t_{2} = 0.031$ eV, $t_{3} = 0.076$ eV, and parameters $\Delta = 0.4$ eV, $\gamma_{0} = 0.02$ eV and 
$\Delta_{\scriptsize{ASO}} = 0.01$ eV are taken from Ref.~\cite{Seibold2017Dec}.
}
	\label{fig:pict}
\end{figure*}

Here, we investigate anomalous Hall and Nernst effects in a two-dimensional electron gas with anisotropic $\mathbf{k}$-cubed Rashba spin-orbit coupling. Anomalous Hall effect has become nowadays one of the most important and commonly used experimental tools, delivering information about magnetization, transport properties, and the system topology~\cite{Nagaosa2010May}. Although the behavior of anomalous Hall conductivity in the presence of linear Rashba coupling has been investigated intensively, the influence of the cubed Rashba coupling on it has got much less attention.  Therefore, the purpose of this paper is to provide the theoretical description of the AHE and ANE in magnetized 2D electron gas with cubic Rashba coupling. Since perovskite oxides have become recently very promising materials for spintronics applications, the effective Hamiltonian derived for 2DEG at LAO/STO interface is considered. However, the presented model and qualitative results may also be applied to other structures that, due to symmetry arguments, can be described by the same Hamiltonian. 

In Section 2 the effective low-energy Hamiltonian describing electronic states of 2DEG with cubic Rashba spin-orbit coupling is described. In this section, the Matsubara-Green's function formalism is applied to the anomalous Hall and Nernst effects. The Berry phase approach is also introduced. The discussion of numerical results is given in Sec. 3. Finally, Sec. 4 contains the final remarks and summary of this work.

\section{Model and method}
\subsection{Effective Hamiltonian}

The electronic energy spectrum describing  STO surfaces and STO/LAO interfaces has been calculated recently within the tight-binding approach and DFT modeling
~\cite{Khalsa2013Jul,Zhong2013Apr,Kim2013Jun,Shanavas2016Jan,Popovic2008Dec,Shanavas2014Oct}.
Based on these calculations the effective Hamiltonian describing the neighborhood of the  $\Gamma$  point in the Brillouin zone has been derived~\cite{Kim2013Jun,Shanavas2016Jan,Zhou2015Jun,Seibold2017Dec}. This energy spectrum is formed by three pairs of bands as presented in Fig.~\ref{fig:pict}(a). These bands are created by $d$-orbitals ($d_{xy}$, $d_{xz}$ and $d_{yz}$) originating mostly from $t_{2g}$ atomic orbitals of Ti. The effective Hamiltonian describing the lowest pair of bands is formed by $d_{xy}$ orbital and has a form of $\mathbf{k}$-linear Rashba Hamiltonian with a negative coupling constant. The middle pair of bands around $\Gamma$ point is described by spin-orbit coupling term which is not only anisotropic in a $\mathbf{k}$-space but also has a cubic dependence on $\mathbf{k}$ (Fig.~\ref{fig:pict}(b)-(e)). The highest, in energy, pair of bands is characterized by effective Hamiltonian with $\mathbf{k}$-linear Dresselhaus-like form of spin-orbit~\cite{Seibold2017Dec}.  
Since electronic transport characteristics in a system with conventional Rashba term is rather well described, within this article, we focus only on the transport properties of quasiparticles from the middle pair of bands. Another word, the aim is to describe transport properties related to quasiparticle states determined by anisotropic $\mathbf{k}$-cubed Rashba spin-orbit interaction. The effective Hamiltonian describing the system under consideration has the following form:

\begin{equation}\label{eq:Hani}
\hat{H} = \frac{\hbar^2 k^2}{2 m}\, \sigma_0 +\lambda \,(k_x^2 - k_y^2)(k_x\, \sigma_y - k_y\, \sigma_x) + M\, \sigma_z ,
\end{equation}
where $k^2 = k_{x}^{2} + k_{y}^{2}$ and $k_{x} = k \cos(\phi)$, $k_{y} = k \sin(\phi)$ are the wavevector components, $m$ is an effective mass of quasiparticle. The Rashba coupling constant is defined as ${\lambda=a^3 \gamma(t_1+t_3-t_2)/\Delta}$, where $a$ is the lattice constant for perovskite oxides, $\gamma$ and $\Delta$ stands for the effective hopping amplitude and energy difference between the $d_{xy}$ orbital and the $d_{xz}$ and $d_{yz}$ orbitals, respectively while  $t_{1,2,3}$ are the tight-binding parameters describing the virtual hopping between $d$-orbital states via $p$-orbitals of the oxygen~\cite{Seibold2017Dec,Khalsa2013Jul,Zhong2013Apr}.
 The last term in the Hamiltonian describes effective exchange interaction with parameter $M$ describing effective magnetization in energy units. The magnetization is oriented in $z$-direction (out-of  2DEG plain) and depends on temperature according to the Bloch relation~\cite{Ashcroft}: $M = M_{0} \left[1-(T/T_{C})^{3/2} \right]$, where $M_0$ is the saturation magnetization, $M_0=M(T=0)$, $T$ is the temperature and $T_C$ denotes Curie temperature.  Here it should be stressed that the above effective Hamiltonian is unbounded from below for large wavevectors what is unphysical. Thus the considerations within this model are restricted only to small quasiparticle densities. Thus, one needs to define the cut-off energy, $E_{c}$ (and corresponding to it cut-off wavevector) below which the states might be occupied and require that chemical potential is far below the cut-off energy, that is $\mu \ll E_{c}$.  Accordingly, for numerical calculations, the module of cut-off wavevector, $k_{c}$, is defined as $k_c=\hbar^2/(3m\lambda)$ and corresponds to the local maximum of the energy dispersion for the lower  (i.e., $E_-$) branch (see  inset in the upper right corner of Fig.~\ref{fig:pict}(f)). Since the cut-off wavevector depends on the effective mass and Rashba coupling constant, thus the energy window related to the reasonable changes of the chemical potential also strongly depend on them. This is shown in Fig.~\ref{fig:pict}(f), where $k_{c}$ is plotted as a~function of the quasiparticle effective mass. Additionally, the Fermi contours fixed for the same Fermi energy, for different values of effective mass are shown. Evidently, the anisotropy of energy bands is more pronounced at higher effective masses. 

The casual Green function corresponding to the Hamiltonian (\ref{eq:Hani}) has the following explicit form:
\begin{equation}
\label{G}
G_{\mathbf{k}}(\varepsilon)  = G_{\mathbf{k} 0}\, \sigma_{0} + G_{\mathbf{k}x}\, \sigma_{x} + G_{\mathbf{k} y}\, \sigma_{y}
\end{equation}
with coefficients:
\begin{eqnarray}
	G_{\mathbf{k} 0} = \frac{1}{2} \left(G_{\mathbf{k} +} + G_{\mathbf{k} -}\right),\hspace{2.8cm}\\
	G_{\mathbf{k} x} =  \frac{\lambda  k^3 }{4 \xi_{\mathbf{k}}} (\sin (\phi )-\sin (3 \phi )) \left(G_{\mathbf{k} +} - G_{\mathbf{k} -}\right), \\
	G_{\mathbf{k} y} =  \frac{\lambda  k^3 }{4 \xi_{\mathbf{k}}} (\sin (\phi )+\sin (3 \phi ))\left(G_{\mathbf{k} +} - G_{\mathbf{k} -}\right),\\
	G_{\mathbf{k} z} = \frac{M}{2 \xi_{\mathbf{k}}} \left(G_{\mathbf{k} +} - G_{\mathbf{k} -}\right),\hspace{2.5cm}
\end{eqnarray}
where\, $G_{\mathbf{k} \pm} = [\varepsilon + \mu - E_{\pm} + i \delta {\mathrm{sgn}}(\varepsilon)]^{-1}$ with the eigenvalues $E_{\pm}=\epsilon_k\pm\,\xi_{\mathbf{k}}$, and $\xi_{\mathbf{k}} = \sqrt{M^2 + \lambda^2 k^6 \cos^2(2\phi)}$. 

\subsection{Anomalous Hall conductivity}
To calculate the anomalous Hall conductivity (AHC), the Matsubara-Green's function formalism has been used (see e.g.~\cite{Mahan2000,Fetter2003,Streda1982Aug,Dyrdal2016Jul}). 
In the linear response regime, the transverse charge current density induced by external electric field can be derived based on the following expression:
\begin{equation}
\label{eq:jx}
j_x(i \omega_{m}) = k_B T \sum_{\mathbf{k},n} \mathrm{Tr} \left\{ \hat{j}_x G_{\mathbf{k}} ( i \varepsilon_n + i \omega_m) \hat{H}_{\mathbf{A}}(i \omega_m) G_{\mathbf{k}}(i \varepsilon_n) \right\},
\end{equation}
where $G_{\mathbf{k}}(i \varepsilon_n)	$ denotes Matsubara-Green's function corresponding to the unperturbed  Hamiltonian (\ref{eq:Hani}) with ${\varepsilon_n = (2n+1)\, \pi\, k_{B} T}$, ${\omega_m = 2 m \pi k_{B} T}$ being Matsubara energies, and $k_B$ is the Boltzmann constant. The charge current density operator is defined as $\hat{j }_i = e\hat{v_i}$ where $e$ is the electron charge and the velocity operator is defined as $\hat{v}_{i} = \hbar^{-1} \partial_{k_{i}} \hat{H}$.  The perturbation Hamiltonian $\hat{H}_{\mathbf{A}}$ describing the coupling of quasiparticles with an external electric field is given in the form:
\begin{equation}
\hat{H}_{\mathbf{A}}(i \omega_{m}) = -\hat{j}_y A_{y}(i \omega_m)\, ,
\end{equation}
where the amplitude of electromagnetic field is linked with the amplitude of electric field through the well known relation: $A_{y}(i \omega_{m}) = -i \hbar E_{y}(i \omega_{m})/i \omega_{m}$.
The sum over Matsubara energies has been done using the method of contour integration  and analytical continuation for Green's function~\cite{Mahan2000}. Finally, the expression for AHC receives the following form:
\begin{eqnarray}
\label{eq:AHC}
\sigma_{xy}(\omega) = - \frac{e^2 \hbar}{\omega} {\mathrm{Tr}} \int \frac{ d^{2} \mathbf{k}}{(2\pi)^{2}} \int \frac{d \varepsilon}{2 \pi} f(\varepsilon)\hspace{2cm}\nonumber\\\hspace{2cm} \times
\left[  \hat{v}_{x} G_{\mathbf{k}}^{R}(\varepsilon + \omega) \hat{v}_{y} [G_{\mathbf{k}}^{R}(\varepsilon) - G_{\mathbf{k}}^{A}(\varepsilon)]\right.  \nonumber\\
+ \left. \hat{v}_{x} [G_{\mathbf{k}}^{R}(\varepsilon) - G_{\mathbf{k}}^{A}(\varepsilon)] \hat{v}_{y} G_{\mathbf{k}}^{A}(\varepsilon - \omega)\right] ,
\end{eqnarray}
where $G_{\mathbf{k}}^{R/A} = \left[(\varepsilon + \mu \pm i \Gamma) \sigma_{0} - \hat{H}\right]^{-1}\nonumber$ stands for retarded/advanced Green's function, $\Gamma$ is a quasiparticle relaxation rate  ($\Gamma=\hbar/2\tau$, $\tau$~--~relaxation time), $\mu$ is a~chemical potential, and $f(\varepsilon)$ denotes the Fermi-Dirac distribution function.
The equation (\ref{eq:AHC}) is a starting formula for further numerical and analytical calculations.

\begin{figure*}
	\centering
	\includegraphics[width=0.9\textwidth]{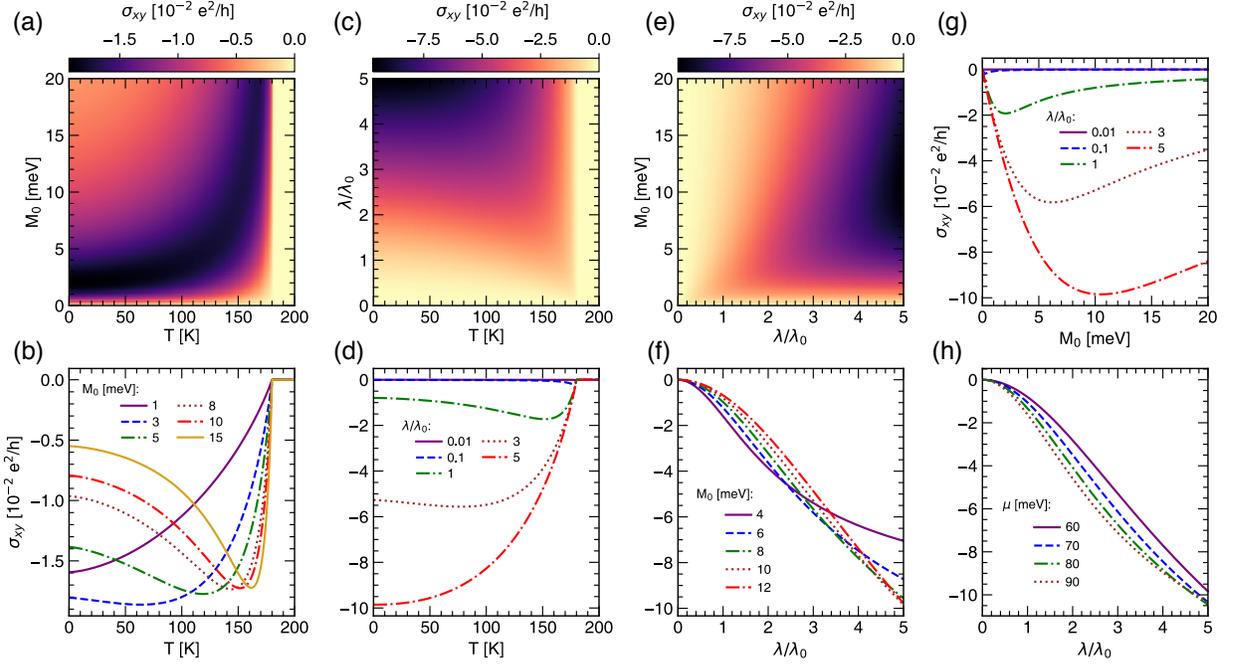}
	\caption{Anomalous Hall conductivity as a function of temperature, $T$, and saturation magnetization, $M_0$, (a), (b), (g);
temperature and Rashba coupling constant, $\lambda$, (c), (d); Rashba constant and saturation magnetization (e), (f), (h);
Other parameters (unless otherwise specified): $\mu = 60$ meV, $M_0 = 10$ meV, $\lambda_0 = 1.07\cdot10^{-30}$ eVm$^3$,
$T = 10$ K, $T_C = 180$ K, $\Gamma = 0.005$ meV, $m = 1.14 m_0$.
	}
	\label{fig:AHE1}
\end{figure*}

\subsection{Anomalous Nernst conductivity}
The anomalous Nernst conductivity (ANC) can be also found based on Matsubara-Green's function formalism. One can start from equation similar to Eq.(\ref{eq:jx}), but with the perturbation Hamiltonian defining as follows:
\begin{equation}
\hat{H}_{\mathbf{{ {\bsfA}}}}(i \omega_{m}) = - \hat{j}_{y}^{h}\, \mathcal{A}_{y}(i \omega_{m}) ,
\end{equation}
where $\hat{j}_{y}^{h}$ is a heat current density operator and  $\bsfA$  is an artificial gravitational vector potential amplitude related to the temperature gradient by the following expression: ${\bsfA(i \omega_{m}) = i \hbar \nabla_{y} T(i\omega_{m})/(i \omega_{m} T)}$ (for details see e.g.~\cite{Luttinger1964Sep,Dyrdal2013Jun,Tatara2015May,Dyrdal2016Jul}).

In turn, it is also known that some thermal transport coefficients obtained within the Kubo-like formalism behave
unphysically when the temperature tends to zero. Thus, to satisfy the Onsager relations, the magnetization currents should be taken into account. Other words, for the anomalous Nernst effect, to obtain results satisfying the third thermodynamic law,  one should add to the expression derived from the Kubo formula an additional term related to the orbital magnetization current density. 
In this manuscript, quite tedious calculations of the orbital magnetization are omitted due to the fact that in the model under consideration the AHE is determined by the topological component. In such a case it is easier to calculate the transverse heat current conductivity, $\beta_{xy}$, which intrinsic contribution is expressed by the entropy density of the electron gas, $\mathcal{S}_{n}(\mathbf{k})$, and the Berry curvature, $\mathcal{B}_{n}^{z}$. Since the transverse heat current conductivity is related to the transverse heat conductivity by the Onsager relation, $\beta_{xy} = \alpha_{xy} T$, ANC is given by the following expression~\cite{Xiao2006Jul,Zhang2008Nov}:
\begin{equation}\label{eq:ANC}
\alpha_{xy} = \frac{e k_B}{\hbar}  \sum_{n=\pm} \int \frac{d^{2} \mathbf{k}}{(2\pi)^{2}} \mathcal{B}_{n}^{z}(\mathbf{k}) \mathcal{S}_{n}(\mathbf{k}) .
\end{equation}
The entropy density for the $n$-th subband is given by the equation:
\begin{equation}
\mathcal{S}_{n}(\mathbf{k}) = - f_{\mathbf{k}}(E_{n}) \ln[f_{\mathbf{k}}(E_{n})] - (1- f_{\mathbf{k}}(E_{n})) \ln[1- f_{\mathbf{k}}(E_{n})]
\end{equation}
and the Berry curvature is calculated from the expression:
\begin{eqnarray}
\mathcal{B}_{n}^{z}(\mathbf{k}) = i \,\nabla_{\mathbf{k}} \times \langle \Psi_{n}| \nabla_{\mathbf{k}} | \Psi_{n} \rangle ,
\end{eqnarray}
with $\Psi_n$ standing for the eigenvector related to the $n$-th eigenvalue of the Hamiltonian (\ref{eq:Hani}).

\begin{figure}
\includegraphics[width=0.93\columnwidth]{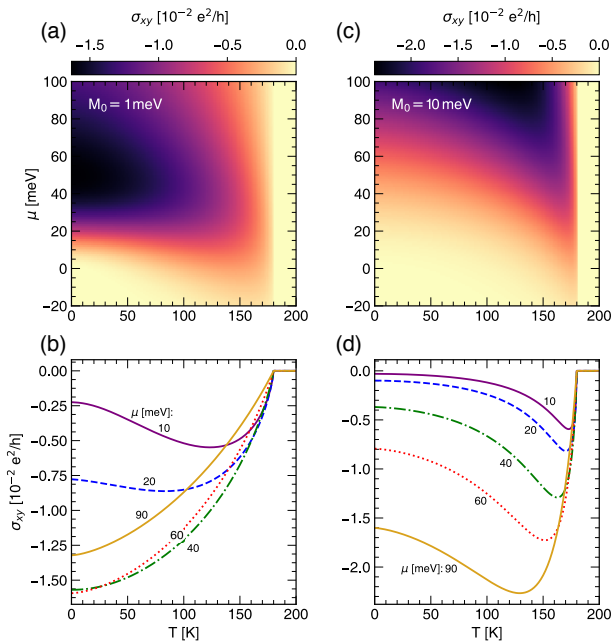}
\caption{Anomalous Hall conductivity as a function of chemical potential, $\mu$, and temperature $T$ (a),(c). The cross sections of the plots (a) and (c) at fixed values of $\mu$  (b), (d).
Other parameters are the same as in Fig.~2.}
	\label{fig:AHE2}
\end{figure}

\section{Results and discussion}
Evaluation of the Eq.~(\ref{eq:AHC}) for AHC and Eq.~(\ref{eq:ANC}) for ANC allows to obtain analytical and numerical results.  Eq.~(\ref{eq:AHC}) has to be integrated analytically over $\varepsilon$ and next the dc-limit ($\omega \rightarrow 0$) has to be taken.  Moreover, after a long discussion about different origins of AHE and proper nomenclature related to its different origins~(see, e.g., {
\cite{Nagaosa2010May}), the AHC is commonly  expressed as a sum of two components:\begin{equation}
\sigma_{xy} = \sigma_{xy}^{I} + \sigma_{xy}^{II}. 
\end{equation}
The first component, $\sigma_{xy}^{I}\,[f'(E_\pm)]$, is the contribution from the states at the Fermi level and the second one, $\sigma_{xy}^{II}\,[f(E_\pm)]$, describes contribution from all states below the Fermi level (so-called Fermi sea or topological component). 


%
%

 Here we consider the quasi-ballistic limit, that means low impurities concentration and weak scattering on impurities, which results in  $\Gamma \rightarrow 0$. In this case, we found that the component $\sigma_{xy}^{I}$ is a few orders of magnitude smaller than the contribution from $\sigma_{xy}^{II}$ and can be neglected. Thus, the electronic properties of the system described by anisotropic $\mathbf{k}$-cubed Rashba model is determined by the  quasi-particle states from the Fermi sea, and:
 \begin{eqnarray}
\label{sigII}
	\sigma_{xy}^{II}= -\frac{3 e^2 M \lambda^2 }{8\pi^2 \hbar} \int d\phi d k\, \frac{k^{5}}{\xi_{\mathbf{k}}^{3}} \cos^2(2\phi)\, \left[ f(E_{-})-f(E_{+})\right].
\end{eqnarray}
This result might be verified easily taking into account the fact that the topological contribution to the AHC may be derived based on the knowledge of the local value of the Berry phase in the system~\cite{Thouless1982Aug,Haldane2004Nov,Xiao2010Jul}:
\begin{eqnarray}
\label{16}
\sigma_{xy}^{II}= - \frac{e^{2}}{\hbar} \sum_{n=\pm} \int \frac{d^{2} \mathbf{k}}{(2\pi)^{2}} \, \mathcal{B}_{n}^{z}(\mathbf{k}) f(E_{n}) . \label{eq:AHCani}
\end{eqnarray}

The Berry curvature for the considered model has the following explicit form:
\begin{equation}
\label{17}
\mathcal{B}_{\pm}^{z}(\mathbf{k}) = \mp \frac{3}{2 \xi_{\mathbf{k}^{3}}} k^4 M \lambda^2 \cos^2(2\phi) .
\end{equation}
Thus, inserting (\ref{17}) into (\ref{16}) gives immediately Eq.~(\ref{sigII}). In turn, taking into account Eq.~(\ref{eq:ANC}) the expression for ANC reads:
\begin{equation}
\alpha_{xy} = \frac{3 e M \lambda^2 k_B}{8\pi^2 \hbar}  \int d\phi d k \, 
\frac{k^5 \cos^2(2\phi)}{\xi_{\mathbf{k}}^{3}}
\left[  \mathcal{S}_{-}(\mathbf{k}) - \mathcal{S}_{+}(\mathbf{k}) \right] .
\end{equation}

\begin{figure*}
	\centering
	\includegraphics[width=0.9\textwidth]{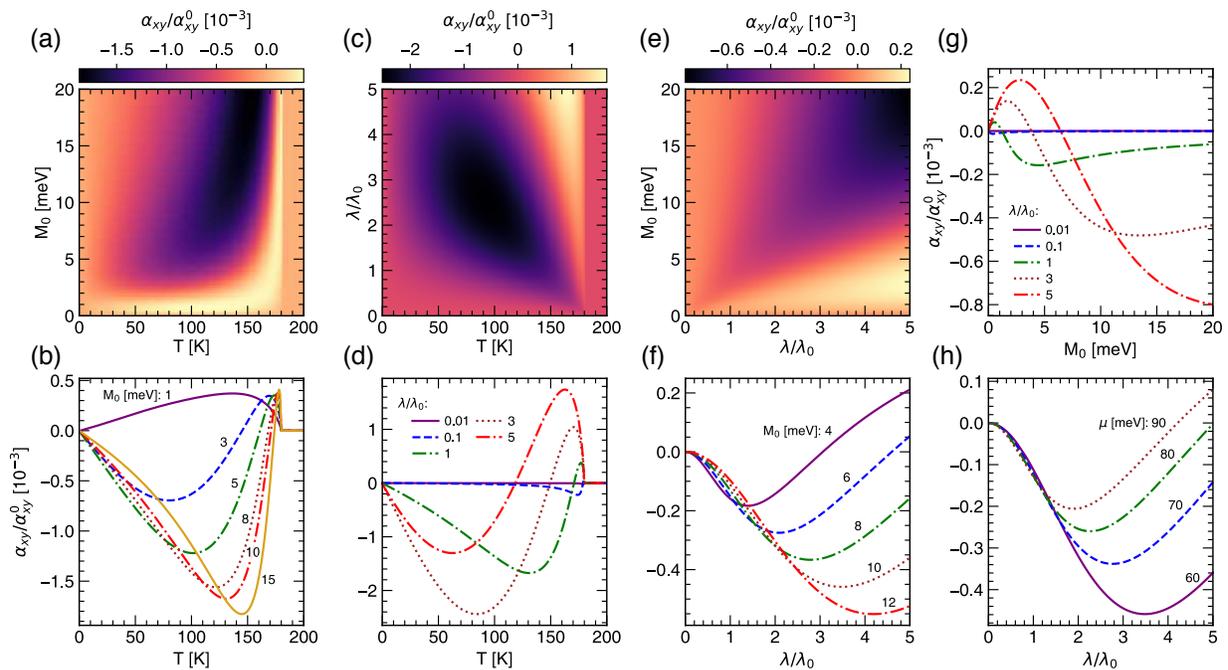}
	\caption{Anomalous Nernst conductivity as a function of temperature, $T$ , and saturation magnetization, $M_0$ , (a), (b), (g); temperature and Rashba coupling constant, $\alpha$, (c), (d); Rashba constant and saturation magnetization (e), (f), (h). Other parameters are as in the Fig.~2.}
	\label{fig:ANE1}
\end{figure*}

Figure~\ref{fig:AHE1} presents numerical results for anomalous Hall conductivity as a function of temperature, $T$, saturation magnetization, $M_{0}$, and Rashba coupling constant, $\lambda$. The AHC  increases slightly with the temperature reaching a maximum at certain value of $T$, and next decreasing to vanish at $T$ equal the Curie temperature ($T = T_{C}$), where the phase transition occurs, and a system becomes nonmagnetic (Figs.~\ref{fig:AHE1} (a)-(d)). Moreover, one can easily see that the maximum value of AHC is shifted to higher temperatures with increasing the saturation magnetization (Figs.~\ref{fig:AHE1} (a),(b)).  For higher temperatures and larger $M_{0}$ the maximum of AHC is well pronounced and precedes the magnetic phase transition.  Furthermore, the competition between the strength of spin-orbit coupling and saturation magnetization is clearly seen in Figs. 2 (e)-(h). AHC increases with the saturation magnetization, reaches a maximum, and then saturates. In general, for the fixed $M_{0}$ and $\mu$, the absolute value of the anomalous Hall conductivity increases with increasing $\lambda$. Note that the chemical potential is fixed with changing the temperature, that is, the number of quasiparticles may be changed. With increasing the temperature the magnetization decreases and the subbands splitting degeneracy also decreases. Moreover, the blurring of quasiparticles distribution also increases. In consequence, the AHC decreases with temperature. The AHC behave quite non-monotonous with the variation of the chemical potential. For the fixed value of saturation magnetization, we observe that AHC increases with increasing the chemical potential but after reaching maximum it decreases and becomes zero at $T=T_{C}$ (Fig.~\ref{fig:AHE2} (a), (b)). Moreover, the maximum of the absolute value of AHC  moves to higher values of $\mu$ and $T$ if the saturation magnetization, $M_{0}$, in the system is higher (compare Fig.~\ref{fig:AHE2} (a), (b) with Fig.~\ref{fig:AHE2}(c), (d)).  Note also that the AHC is almost zero when only one subband is occupied (that is, for Fermi level in the Zeeman gap). Only when the temperature is sufficiently large, in comparison to $M_{0}$, the thermal smearing of charge carriers distribution in both bands leads to nonzero AHC, also for negative values of $\mu$, what is seen in Fig.~\ref{fig:AHE2}. 

Figure~\ref{fig:ANE1} shows the anomalous Nernst conductivity as a function of the same parameters as for AHC, that is the temperature, $T$, saturation magnetization, $M_{0}$, and Rashba constant, $\lambda$. For fixed value of chemical potential (here $\mu = 60$~meV), and sufficiently large saturation magnetization one can observe that the  anomalous Nernst conductivity increases almost linearly with temperature and then, for a certain value of $T$, decreases abruptly and change the sign to reach well define pick for temperatures preceding the Curie temperature (Figs.~\ref{fig:ANE1} (a)-(d)). Both maxima (maximal negative and positive values) occur for higher temperatures when saturation magnetization, $M_{0}$ is higher (Fig.~\ref{fig:ANE1}(b)) or the spin-orbit coupling parameter is smaller, $\lambda$ (Fig.~\ref{fig:ANE1}(d)).  Moreover, one can see very non-monotonous behavior of ANC as a function of both $M_{0}$ and $\lambda$ (Figs.~\ref{fig:ANE1} (e)-(h)). When the spin-orbit coupling is sufficiently large (i.e., it dominates the exchange interaction), ANC takes positive values and increases with $M_{0}$. After reaching a maximum, the ANC decreases, and for a certain value of $M_{0}$ it changes sign and approaches minimum. Finally, the absolute value of  ANC slightly decreases and saturates for sufficiently large values of $M_{0}$. When exchange interaction dominates the  Rashba one, the ANC is always negative. This is also very good seen in Fig.~\ref{fig:ANE2} where ANC is presented as a function of temperature and chemical potential for the two different values of $M_{0}$. 

One of the most important features in the behavior of ANC is the change of its sign, that occurs for temperatures preceding the magnetic phase transition. The similar sign reversal of ANC as a function of temperature has been observed experimentally in different oxides materials (such as LSMO thin layers and SRO crystals)~\cite{Bui2014Sep,Miyasato2007Aug} as well as in ferromagnetic semiconductors~\cite{Pu2008Sep}. In the case of all these experimental data, the intrinsic mechanism has been confirmed as a dominant one. This behavior seems to be in agreement with our theoretical studies that show unambiguously that the topological contribution governs the behavior of both anomalous Hall and anomalous Nernst effect. Moreover, this is also consistent with our previous study for the two-dimensional gas with isotropic cubic Rashba coupling~\cite{Krzyzewska2018Oct}.

\begin{figure}
	\includegraphics[width=0.93\columnwidth]{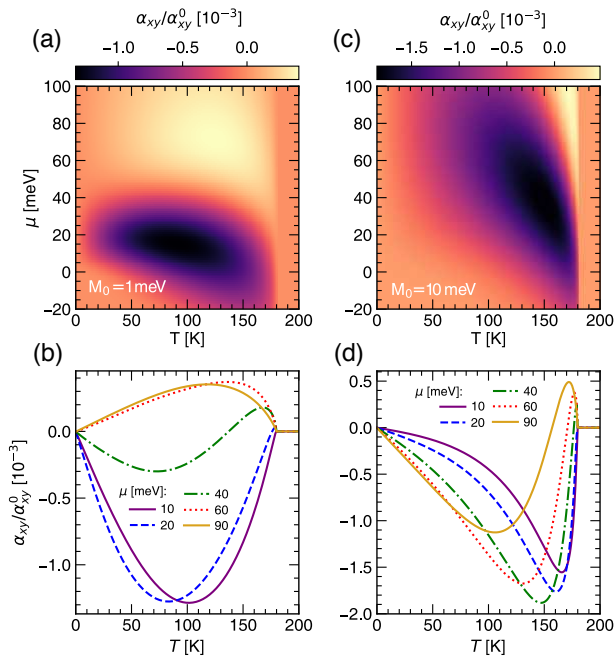}
	\caption{ANC as a function of chemical potential, $\mu$, and temperature, $T$  (a),~(c). The cross sections of the plots (a) and (c) at fixed values of $\mu$ (b),~(d).
Other parameters are the same as in Fig.~2.}
	\label{fig:ANE2}
\end{figure}

\section{Conclusions}
The anomalous Hall and Nernst effect have been studied in the magnetized two-dimensional electron gas with anisotropic $\mathbf{k}$-cubed Rashba spin-orbit interaction. It has been shown that the topological term determines anomalous Hall and anomalous Nernst conductivity. The contribution from the states at the Fermi level to both conductivities are few orders of magnitude smaller and does not affect the total system responses. Such behavior of AHE (ANE) in the systems revealing $\mathbf{k}$-cubic Rashba interaction is distinct in comparison to the systems with $\mathbf{k}$-linear Rashba coupling where AHE is nonzero only when the carriers relaxation times are finite and spin-dependent (see, e.g.,~\cite{Inoue2006Jul}). Moreover, the change of sign in ANC has been observed in temperature dependences. The sign reversal precedes the magnetic phase transition and has been observed recently in experiments for magnetic perovskite oxides. 
The system responses for the model studied in this paper might be changed when one takes into account scattering processes (taking corrections related to the vertex correction, skew-scattering, and side-jump, randomness of spin-orbit coupling, etc.) related to the impurities with a magnetic moment and spin-orbit coupling. All these processes (that will be studied separately elsewhere) may modify the contribution from the states at the Fermi level remaining the topological contribution unchanged.

\section{Acknowledgments}
We thank Prof. J. Barna\'{s} and Prof. J. Berakdar for fruitful discussions. 
A. Dyrda\l{}  acknowledges the support of German Research Foundation (DFG) through SFB 726.

%










\end{document}